\documentclass[12pt]{iopart}

\usepackage{graphicx}
\begin{document}

\title[QD and LQU in a vertical quantum dot]{Comparison of quantum discord and local quantum uncertainty in a vertical quantum dot}

\author{E. Faizi, H. Eftekhari}

\address{Physics Department, Azarbaijan shahid madani university}
\ead{ efaizi@azaruniv.edu}
\ead{ h.eftekhari@azaruniv.edu}
\begin{abstract}
In this paper, we consider quantum correlations (quantum discord and local quantum uncertainty) in a vertical quantum dot. Their dependencies on magnetic field and temperature are presented in detail. It is noticeable that, quantum discord and local quantum uncertainty behavior is similar to a large extent. In addition, the time evolution of quantum discord and local quantum uncertainty under dephasing and amplitude damping channels is investigated. It has been found that, for some Bell-diagonal states quantum discord is invariant
under some decoherence in a finite time interval [Phys. Rev. Lett. 104, 200401 (2010)]. Also, our results show that quantum discord is invariant under dephasing channel for a finite time interval in a vertical quantum dot, while this phenomenon does not occurs for local quantum uncertainty case.

\end{abstract}


\section{Introduction}
The quantum correlation for a quantum state contain entanglement and other type of nonclassical correlations. It is known that the quantum correlations are
more comprehensive than entanglement \cite{C. H. Bennett, W. H. Zurek}. A prominent and widely approved quantity of quantum correlation is the quantum discord (QD) \cite{H. Ollivier, K. Modi} which indicates the quantumness of correlations.
Considerable progresses have been made about the significance and applications of quantum
discord. Particularly, there are strict expressions for quantum discord for some two- qubit states, like as
for the X states \cite{M. Ali, B. Li, Q. Chen, M. Shi}. Despite of quantum discord, a lot of other measures of quantum correlation have been given, such as the GMQD \cite{S. Luo, B. Dakic}, MID \cite{S . Luo} and quantum deficit \cite{J. Oppenheim, M. Horodecki}. Lately Girolami et. al. \cite{D. Girolami} proposed the concept of local quantum uncertainty which determines the uncertainty in a quantum
state as a result of measurement of a local observable. However, such quantifier is strong criterion to be considered as a accurate measure of quantumness in quantum states. Although because of inherent optimization, finding explicit
expression is a difficult problem for most of the quantum correlations measures. For instance, the value of quantum discord is not
 known even for general bipartite qubit system. In bipartite systems with higher dimension, the results are known for
only some certain states. Nevertheless, local quantum uncertainty (LQU) has closed form only for any qubit-qudit system.\\
 Quantum dot (as the artificial atoms) devices are a well- controlled object for studying quantum many- body physics. Also, ground state single exciton qubits in quantum dots have been introduced for quantum computation tasks \cite{P. Solinas}. So it is worthwhile, investigation of the characteristics and properties of the quantum dot.\\
Decoherence of the quantum system due to interacting with its surrounding is the important difficulty to perform quantum computation tasks. Therefore, it is inevitable to specify the dynamical properties of quantum correlations for preserve the protocol to against decoherence.  Many investigation have been paid to dynamics of quantum correlations both theoretically and experimentally in the Markovian \cite{M. Piani, J. Maziero} and non- Markovian \cite{Fanchini} environment. For instance,  there is an
many investigation on decoherence due to spin environment \cite{A. Hutton, F. M. Cucchietti, D. Rossini, H. T. Quan}, like single qubit coupled to the environment and two qubits coupled to the environment.

 In this paper, our goal is to study QD and LQU in a vertical quantum dot. Their dependencies on magnetic field and temperature are also investigated.\\
The parer is organized as follows. In sec. 2, we recall QD, LQU briefly. In sec. 3 we will investigate these quantities in vertical quantum dot and give a detailed comparison. The effect of magnetic field and temperature are illustrated. Sec. 4 is devoted to  the dynamics
of QD in dephasing and amplitude damping model, and the dynamics of LQU is compared with that of QD. In the last section, the conclusions are given.

\section{Quantum discord, Local quantum uncertainty}
\subsection{QD}
For a bipartite quantum system, the quantum mutual information between the two subsystems A and B is as follows:
\begin{eqnarray} I({\rho}_ {AB})=S({\rho}_A)+S({\rho}_B)-S({\rho}_{AB}),
\end{eqnarray}
where $S({\rho})=-Tr(\rho\log_2\rho)$ is Von- Neumann entropy of the density matrix $\rho$. The quantum mutual information has fundamental physical importance, and is generally applied as a measure of total correlations that contain quantum and classical ones. The classical correlation may be defined by projective measurement. Assume one carry out a set of projective measurements $\Pi_k^B$ on the subsystem B, then the probability of measurement with outcome k is $P_k =Tr_{AB}[({I^A}\otimes{\Pi_k^B}){\rho}_ {AB}({I^A}\otimes{\Pi_k^B})]$ where $I^A$ the identity operator for subsystem A. After
this measurement, the state of subsystem A is characterized by the conditional density operator  ${\rho}_ {A\mid{B}}=Tr_B[({I^A}\otimes{\Pi_k^B}){\rho}_ {AB}({I^A}\otimes{\Pi_k^B})]/P_k$.  We determine the upper bound of the difference between the Von- Neumann entropy $S(\rho_A)$ and the based on measurement quantum conditional entropy $\sum_kP_kS({\rho}_ {A\mid{k}})$ of subsystem A, i.e. \cite{H. Ollivier, S . Luo,  J. Maziero, V. Vedral},
\begin{eqnarray} C({\rho}_{AB})=sup_{\{\Pi_k^B\}}[S({\rho}_A)-\sum_kP_kS({\rho}_ {A\mid{k}})],
\end{eqnarray}
as the classical correlation of the two subsystems. The maximum is taken for whole probable types of projective measurements. Finally, the quantum discord is specified as the difference between the total and classical correlations \cite{H. Ollivier, J. Maziero, V. Vedral}.
\begin{eqnarray} D({\rho}_ {AB})=I({\rho}_ {AB})-C({\rho}_ {AB}),
\end{eqnarray}

\subsection{LQU}
Lately, a measure of quantum correlations for bipartite quantum systems namely the local quantum uncertainty (LQU) is introduced by D. Girolami \cite{D. Girolami}. The LQU is defined as follows:
\begin{eqnarray} U_A=\min_{K^A}I(\rho_{AB}, K^A),
\end{eqnarray}
where the two parts denoted by A and B, the minimum is optimized over all of the non degenerate local projective operators on part A:
$K^A=\Lambda{^A}\otimes{{I}_B}$, and
\begin{eqnarray} I(\rho,K)=-\frac{1}
{2}Tr\{[\sqrt{\rho},K^A]^2\},
\end{eqnarray}
is the information which introduced in Ref. \cite{E. P. Wigner}. The closed form of the LQU for $2\times{d}$ quantum systems is \cite{D. Girolami}:
\begin{eqnarray} U_A=1-\lambda_{max}(W),
\end{eqnarray}
 In which $\lambda_{max}$ is the maximum eigenvalue of the $3\times{3}$ matrix W with the elements $W_{ij}=Tr\{\sqrt{\rho}(\sigma_{i}\otimes{I})\sqrt{\rho}(\sigma_{j}\otimes{I})\}$ and $\sigma_i$ $i=1, 2, 3$ is the Pauli matrices.
\section{Quantum discord and Local quantum uncertainty in a vertical quantum dot}
In this section, we will investigate QD and LQU in a vertical quantum dot. The effects of magnetic field and temperature on these outstanding characteristics of quantum physics are demonstrated. Moreover, we will compare these quantities and illustrate their different properties.\\
The reduced Hamiltonian of the quantum dot is written as \cite{L. G. Qin}:
\begin{eqnarray} \hat{H}=\frac{k_0}{4}\hat{S_1}.\hat{S_2}-\gamma{B_0}\hat{S^3},
\end{eqnarray}
Where $\gamma$ is gyromagnetic ratio, $k_0=\delta-2E_s>0$ is the bare value at B = 0 and $\hat{S}^3$ is the third component of total spin. $B_0$ is the magnetic field of the
degenerate point, $\delta$ is the level spacing and $E_s$ is the exchange energy.

 In the standard basis, $\{|00\rangle, |01\rangle, |10\rangle, |11\rangle\}$ , the density matrix $\rho(T)$ of the system reads \cite{L. G. Qin}
 \begin{eqnarray}
\rho(0)={\frac{1}{Z}} \left(
\begin{array}{cccccccccccccccc}
u && 0 && 0 && 0\\
0 && w && y && 0\\
0 && y && w && 0\\
0 && 0 && 0 && v\\
\end{array}
\right).
\end{eqnarray}
In which the nonzero matrix elements are given by
\begin{eqnarray} u=\exp(-\frac{k_0-16\gamma{B_0}}{16T});\nonumber\\
v=\exp(-\frac{k_0+16\gamma{B_0}}{16T});\nonumber\\
w=\frac{1}{2}[\exp(\frac{-k_0}{16T})+\exp(\frac{3k_0}{16T})];\nonumber\\
y=\frac{1}{2}[\exp(\frac{-k_0}{16T})-\exp(\frac{3k_0}{16T})],
\end{eqnarray}
and $Z=u+2w+v$. Here $\gamma$ and $B_0$ always appear in the form $\gamma{B_0}$
and thus we can consider it as $\gamma{B_0}=r$.\\
We can find that, the ground state of the Hamiltonian in eq. (7) become separable when the magnetic field is strong enough.
In addition, the ground state will be entangled when
the $k_0$ is large enough (for more detail see the eigenvalues and eigenvectors of reduced
Hamiltonian Eq. (7) in \cite{L. G. Qin}). From this aspect, we can say that a strong magnetic field will shrink the quantum
correlation measured by the QD, however a large $k_0$ can cause a large amount of quantum correlation.\\
The matrix (8) is the X- matrix whose discord has been studied in \cite{M. Ali}.
Although this reference contains a mistake concerning the number of arbitrary optimization parameters in
the calculation of the classical part of mutual correlations \cite{M. Ali, Y.Huang},
this mistake is not important in our case because the element $\rho_{14}$ is zero in density matrix (8). As a consequence, we have only one optimization
parameter. Thus, we use the algorithm developed
in the above reference for the calculation of discord.
By the density matrix given in eq. (8), we  can find the analytical expressions of QD as given in ref. \cite{M. Ali, F. F. Fanchini}. The QD of the density matrix in eq. (8) can
be computed directly and it takes the following expression:
\begin{eqnarray} D=min\{D_1,D_2\},
\end{eqnarray}
where $D_1$ and $D_2$ read respectively as:

\begin{eqnarray} &D_1=S(\rho_A)-S(\rho)-\frac{1}{Z}[v\log_2{\frac{v}{w+v}}+w\log_2{\frac{w}{w+v}}],\\\nonumber
&-\frac{1}{Z}[u\log_2{\frac{u}{w+u}}+w\log_2{\frac{w}{w+u}}],\\\nonumber
&D_2=S(\rho_A)-S(\rho)-\frac{1-\Gamma}{2}\log_2\frac{1-\Gamma}{2}-\frac{1+\Gamma}{2}\log_2\frac{1+\Gamma}{2}.
\end{eqnarray}
where $\rho_A$  is the reduced density matrix of $\rho$ in eq. (8) by tracing off the second party and $\Gamma$ satisfies the relation
$\Gamma=\sqrt{{(u-v)^2+4|y|^2}}/{Z}$. The  analytical
expression of QD shows that the QD depends on the temperature, the magnetic field and $k_0$.\\
The LQU of the thermal density matrix in eq. (8) takes the expression of

\begin{eqnarray} U_A=1-\max\{\lambda_1,\lambda_2\},
\end{eqnarray}
where $\lambda_1=2(\sqrt{u}+\sqrt{v})(\frac{\sqrt{w-y}}{2}+\frac{\sqrt{w+y}}{2})$ and
 $\lambda_2=(u+v)+2(\frac{\sqrt{w-y}}{2}+\frac{\sqrt{w+y}}{2})^2-2(\frac{\sqrt{w+y}}{2}-\frac{\sqrt{w-y}}{2})^2$ are the eigenvalue of the $3\times{3}$ matrix W.\\
 After calculations, we find that QD and LQU are symmetric under change of r to -r, so we will consider only
$r>0$ in our calculations. The influence of parameters on QD and LQU in quantum dot is discussed in detail as follows.\\

\begin{figure}
\includegraphics[width=3in]{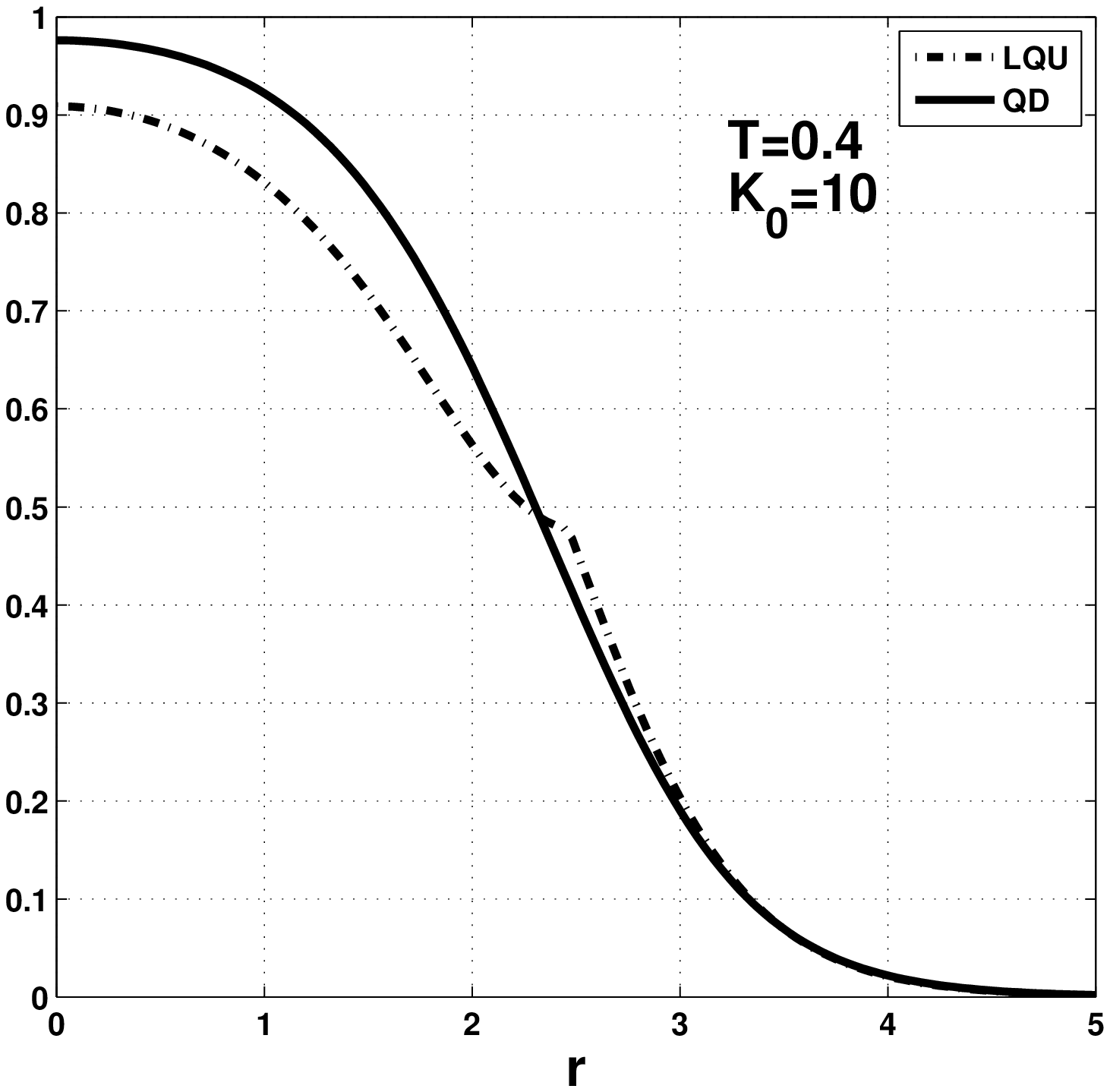}
\includegraphics[width=3in]{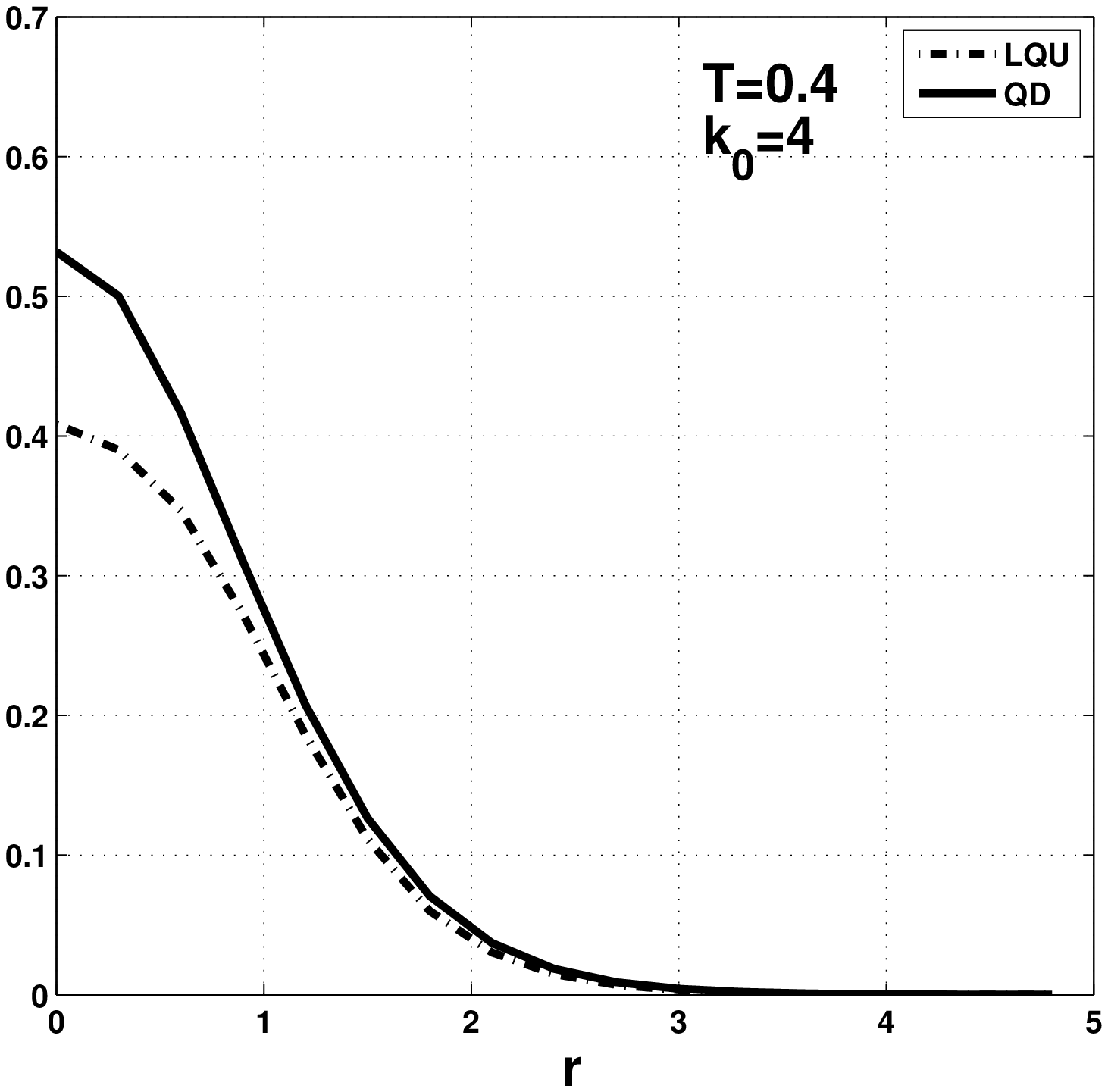}
\caption{Comparison of QD and LQU  vs $r$ and fix $k_0$ and T.}
 \label{fig1}
\end{figure}

\begin{figure}
\includegraphics[width=3in]{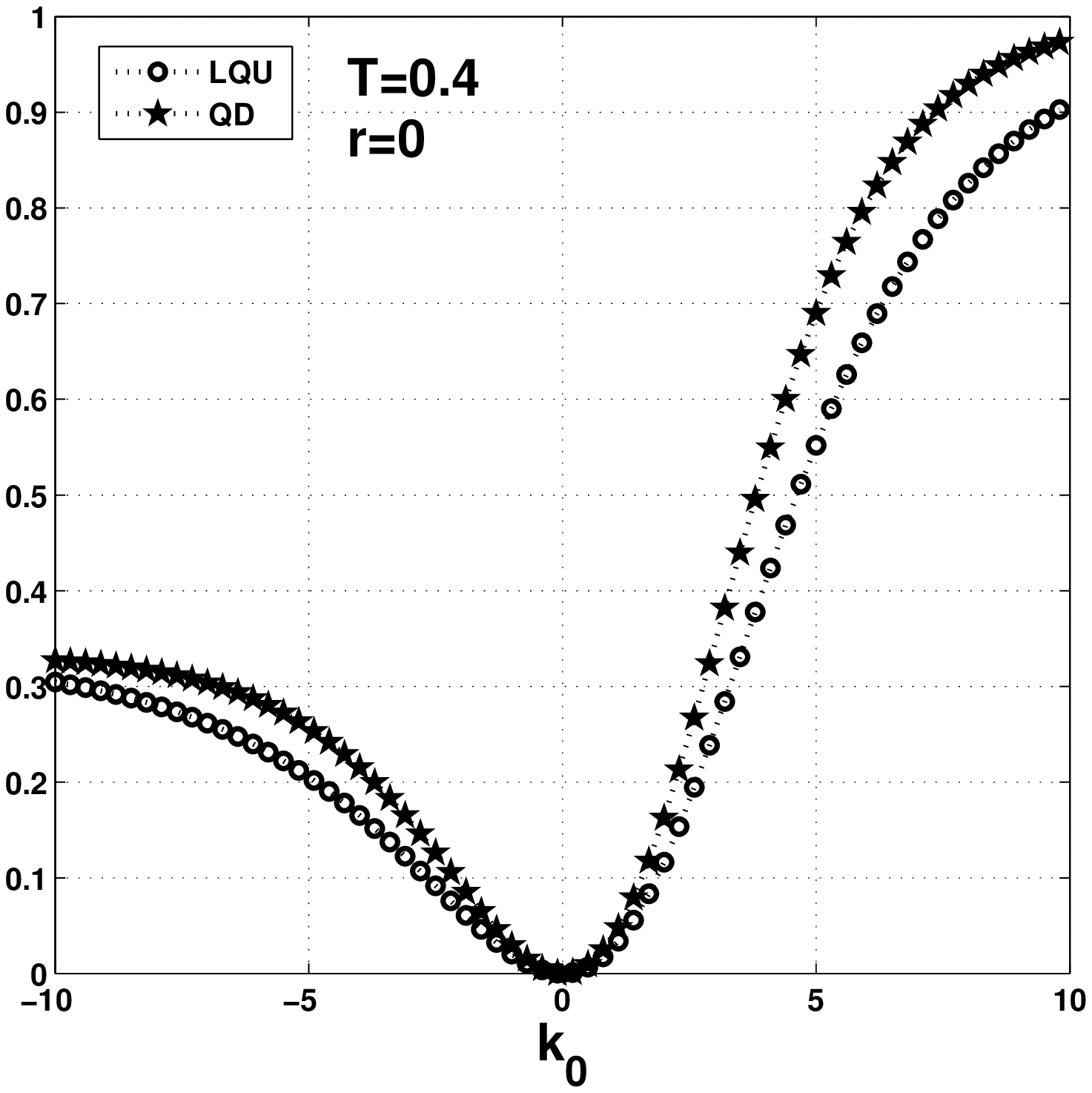}
\includegraphics[width=3in]{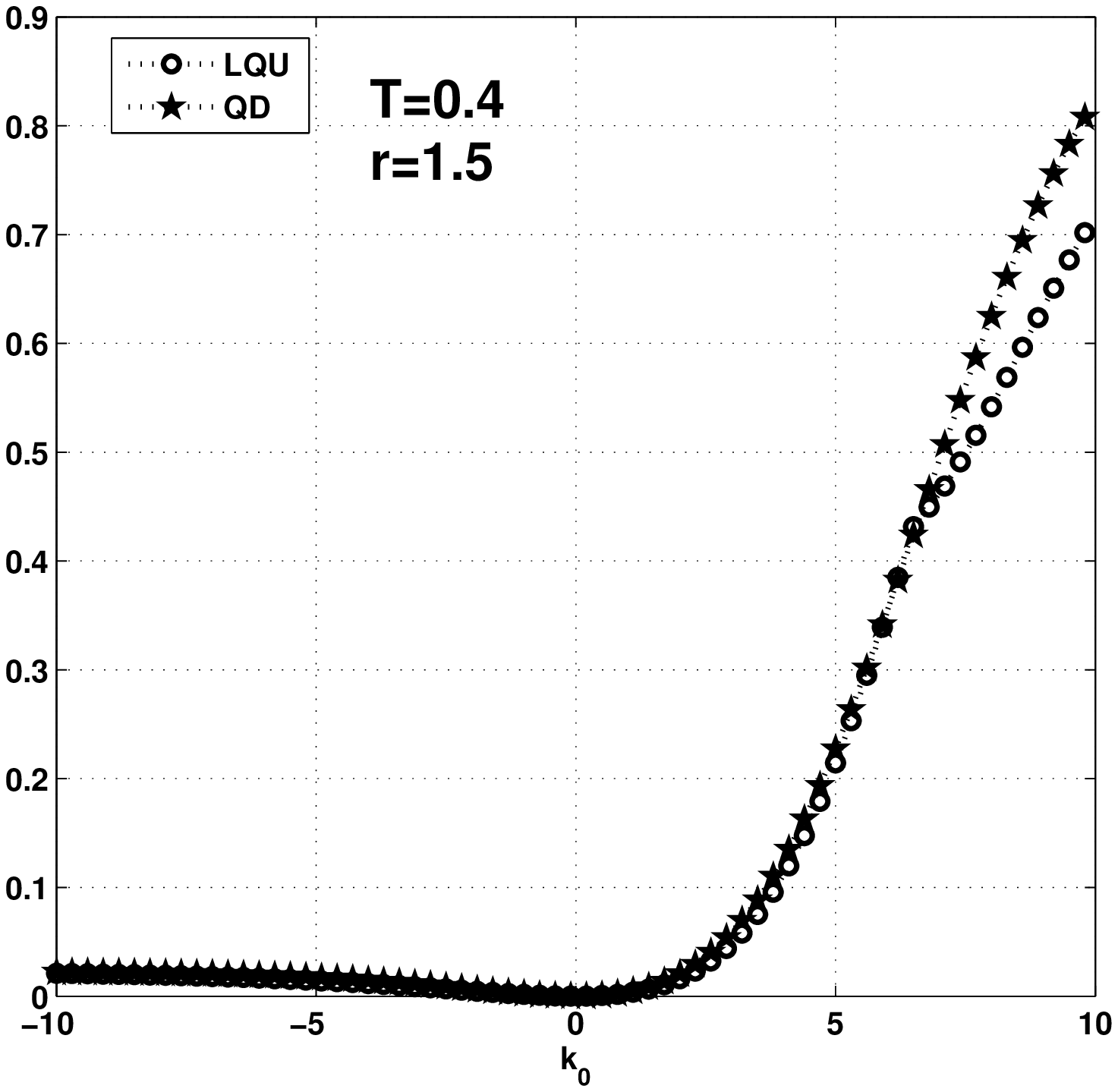}
\caption{Comparison of QD and LQU  vs $k_0$ and fix $r$ and T.}
 \label{fig2}
\end{figure}

\begin{figure}
\includegraphics[width=3in]{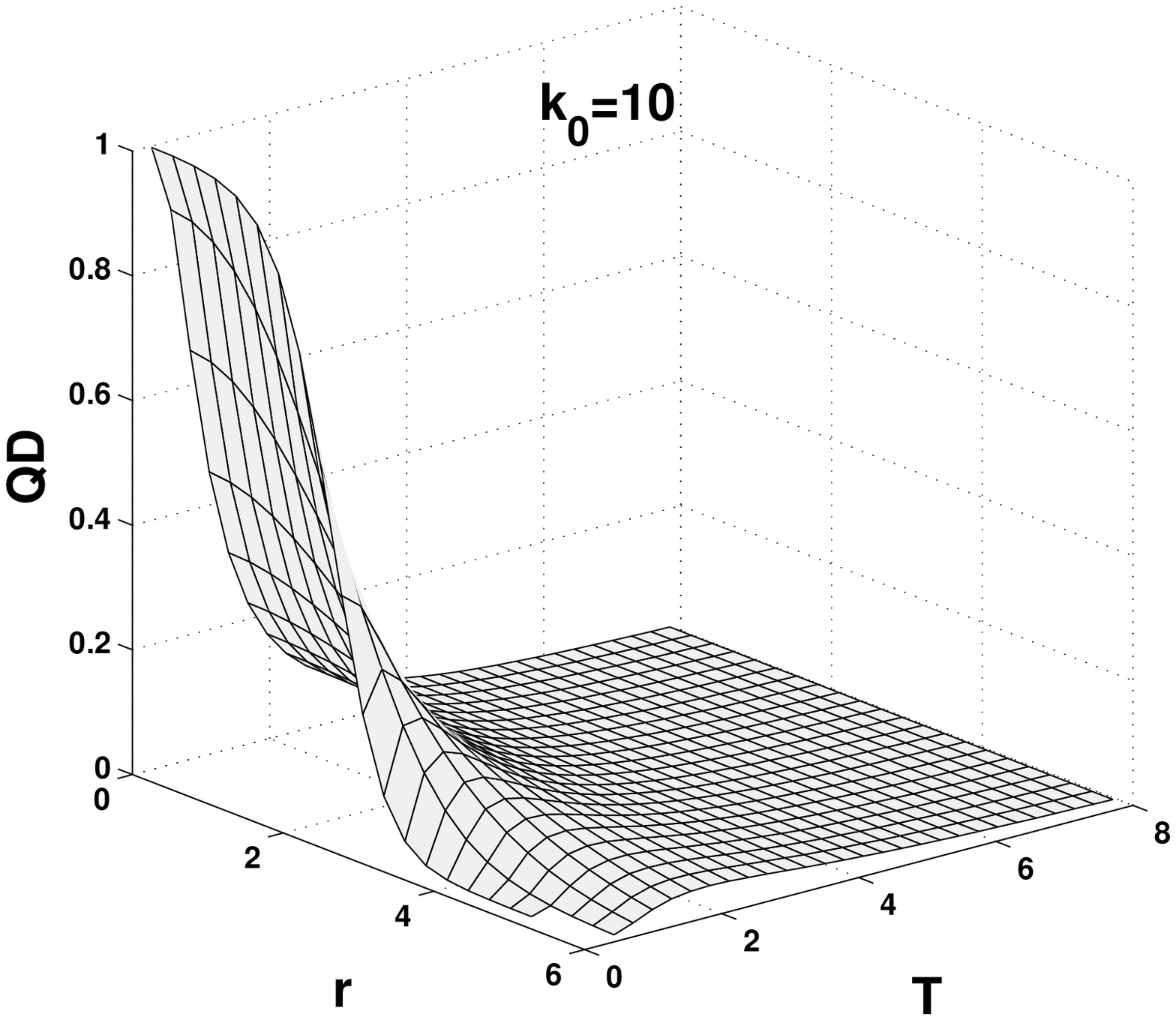}
\includegraphics[width=3in]{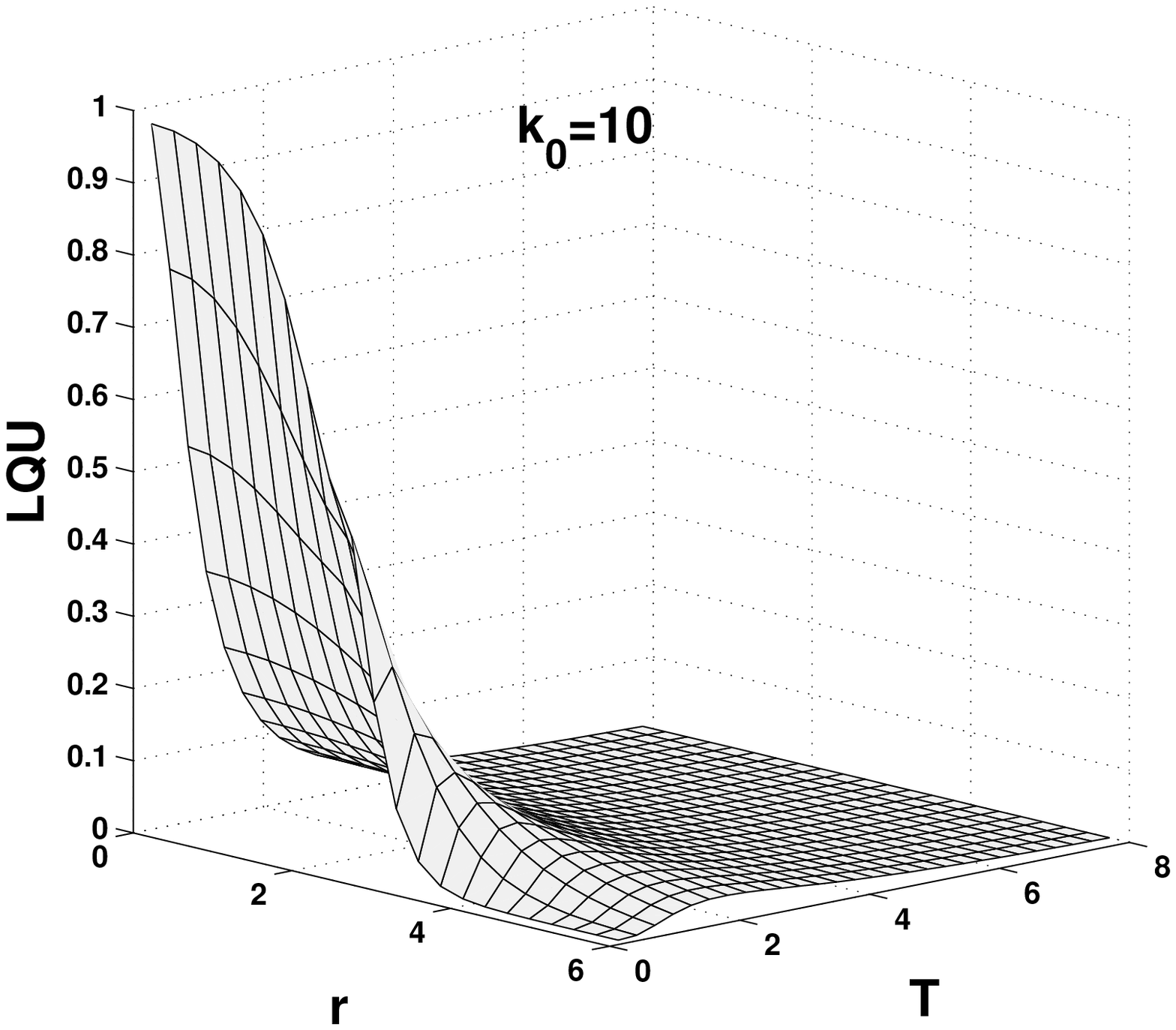}

\caption{QD and LQU vs temperature (T) and r in the case of  $k_0=10$.}
 \label{fig3}
\end{figure}

\begin{figure}
\includegraphics[width=3in]{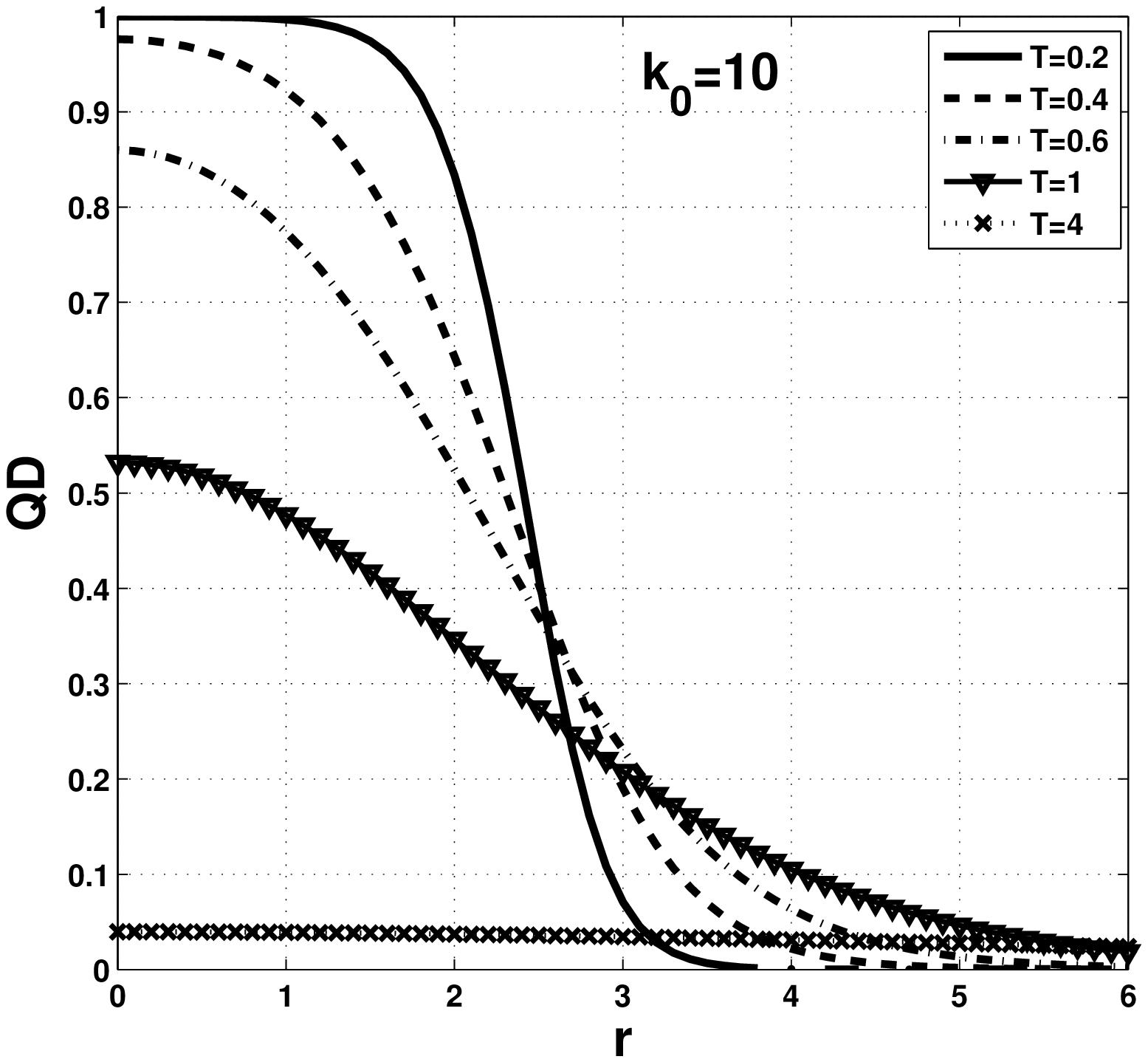}
\includegraphics[width=3in]{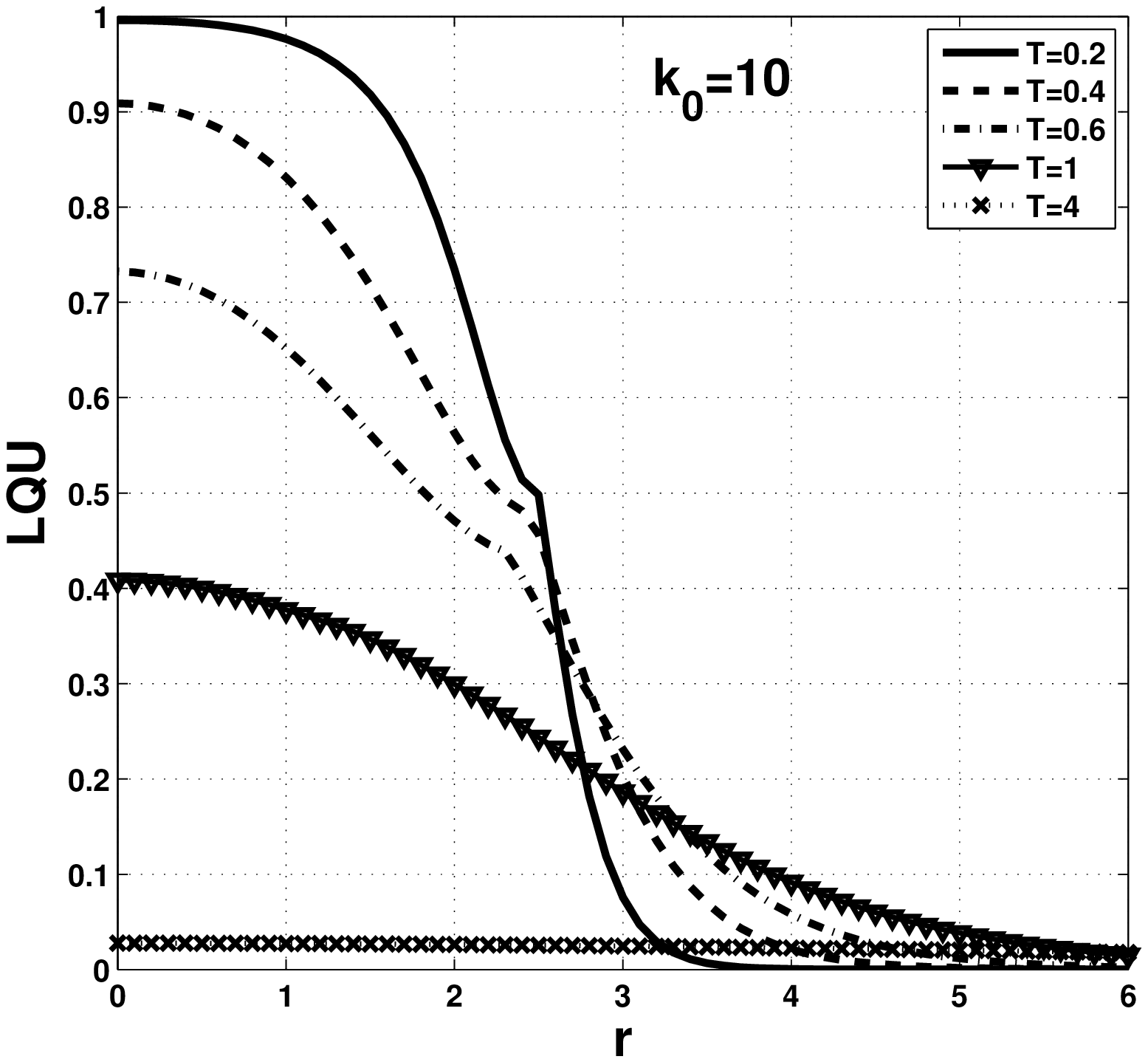}

\caption{QD and LQU vs r for different temperature (T) and $k_0=10$.}
 \label{fig4}
\end{figure}

At first, we analyze the sensitivity of the parameters $k_0$ and r for the QD and the LQU
and the results are given in Figures 1 and 2. From Figures 1 and 2 we can see that the behaviors of QD and LQU are similar to a large extent. When the temperature is not zero, QD and LQU changes with the r when the $k_0$ is fixed. The higher the r is, the smaller the QD is, and the smaller the LQU is. In this sense, we can find that high r can shrink the QD and LQU. Also, both of QD and LQU decreases asymptotically to a very small value. From Figure 2 we can see that $k_0>0$
 show more quantum correlation than $k_0<0$ and large $k_0$ cause large quantum correlation. Although for $k_0<0$, there is no entanglement (see ref. \cite{L. G. Qin}), QD and LQU exist. Moreover, we can see that when r is zero quantum correlations have higher value this is because the
ground state become the maximally entangled state.  \\
Secondly, we examine the effect of the temperature on the QD and the LQU and the results are given in Figure 3. From Figure 3, for the case $T\neq0$, the QD and LQU decreases
by increasing temperature. Nevertheless, they decreases  more slowly when  the temperature is higher. From Figure 3, we can find
that the QD and the LQU is not sensitive to the magnetic field when the temperature takes a value larger than about 2. This point indicates that the
quantum correlation measured by the QD and LQU may not be affected by the magnetic field efficiently when the system  temperature has high value. As to give a better illustration about the sensitivity of the QD and the LQU to the temperature, we plot Figure 4.
From Figure 4, we can find that the QD and LQU is sensitive to the magnetic field when the temperature is low. While for the high temperature
of the case  T=4, they are not sensitive anymore and  remain stable.

\section{Evolution of QD and LQU in the vertical quantum dot under noisy channels}
In order to calculate the quantum discord between two qubits subject to dissipative channels, we consider the following approach. The dynamics of two qubits which interacting independently with distinct environments is described by the solutions of the Born-Markov-Lindblad
equations \cite{H. Carmichael}, that can be  acquired appropriately by the Kraus operator method \cite{M. A. Nielsen}. Given an initial state for
two qubits $\rho(0)$, its time evolution can be written as
\begin{eqnarray} \rho(t)=\Sigma_{\mu,\nu}E_{\mu,\nu}\rho(0){E_{\mu,\nu}^\dag},
\end{eqnarray}
where the Kraus operators $E_{\mu,\nu}=E_\mu\otimes{E_\nu}$ \cite{M. A. Nielsen} satisfy $\Sigma_{\mu,\nu}{E_{\mu,\nu}^\dag}E_{\mu,\nu}=I$ for all t. The operators $E_{\mu}$  characterize the one- qubit quantum channel effects.  We present below what happens to the QD and LQU in
 for two qubit of the dephasing and amplitude damping channels.

\subsection{Dephasing channel}
 Here we examine time evolution of the vertical quantum dot under first phase damping and then amplitude damping channels. We will begin by obtaining the time dependence of QD and LQU for the vertical quantum dot.
Recently, it has been shown that for some Bell- diagonal states (BDS), their quantum discord are invariant
under some decoherence for a finite time interval \cite{L. Mazzola}. An interesting question is that such phenomenon occurs in other
systems?\\
 In the next of this section we consider that the state of density matrix $\rho$ in Eq .(8) undergoes the dephasing channel.
Kraus operators for a dephasing channel given by $E_0=diag(1,\sqrt{1-\gamma})$ and $E_1=diag(0,\sqrt{\gamma})$ where $\gamma=1-e^{-\Gamma{t}}$, $\Gamma$ denoting decay rate \cite{M. A. Nielsen}.
Under the effect of phase noise the only time dependence is in y and other element in density matrix remain unchanged:
\begin{eqnarray} y(t)=y(0)(1-\gamma),
\end{eqnarray}
 In fact, the time- dependent parameter $\gamma$ may be different for qubits A and B,
but we take it identical.\\
The results are shown in figure 5. We can see that the behavior of QD and LQU under the effect of this channel is different. In particular we note the evolution of QD, which its smoothly behavior at a finite time is noticeable (and remains stable). However, LQU by increasing t decrease monolitically.
\begin{figure}
\includegraphics[width=3in]{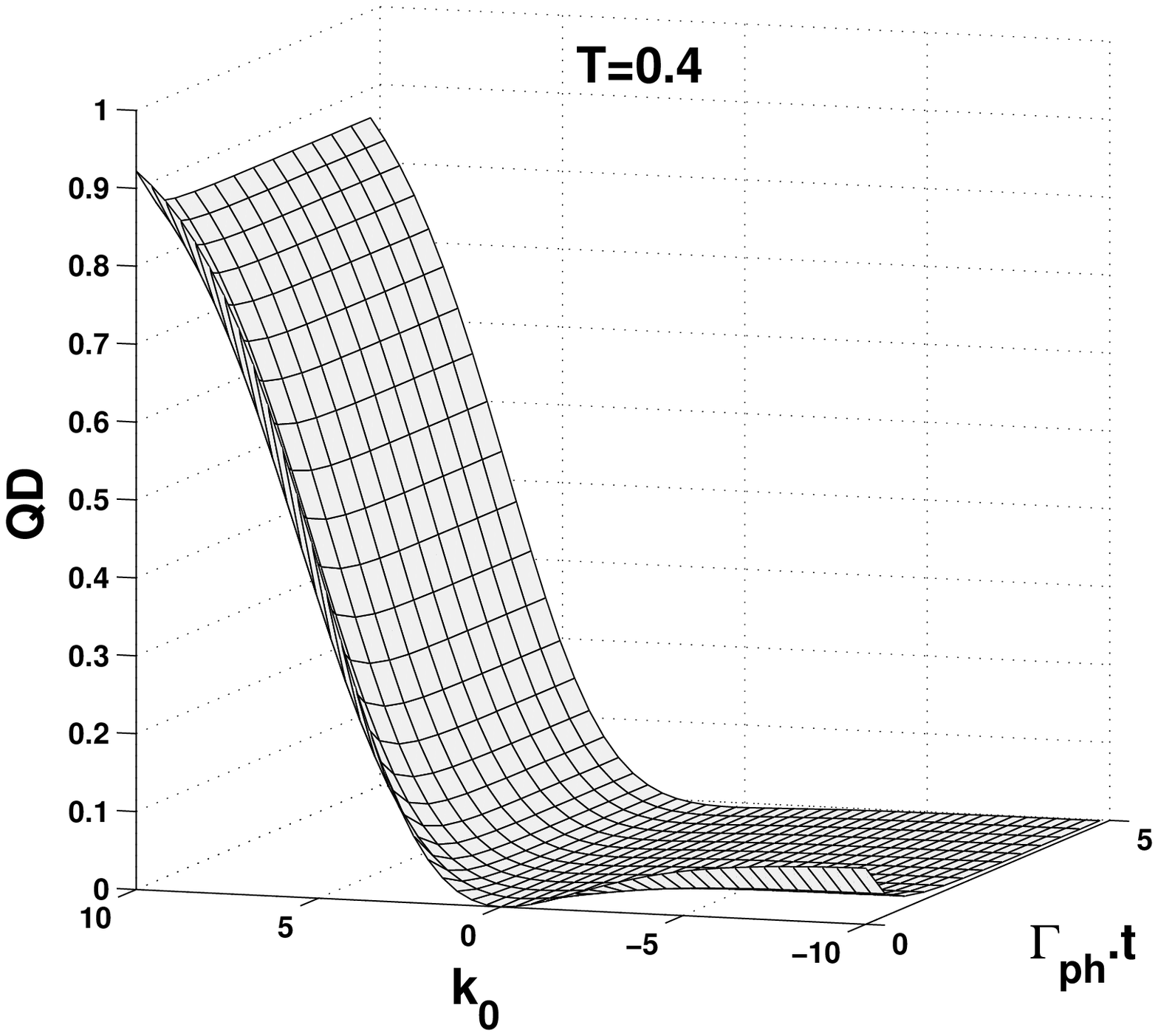}
\includegraphics[width=3in]{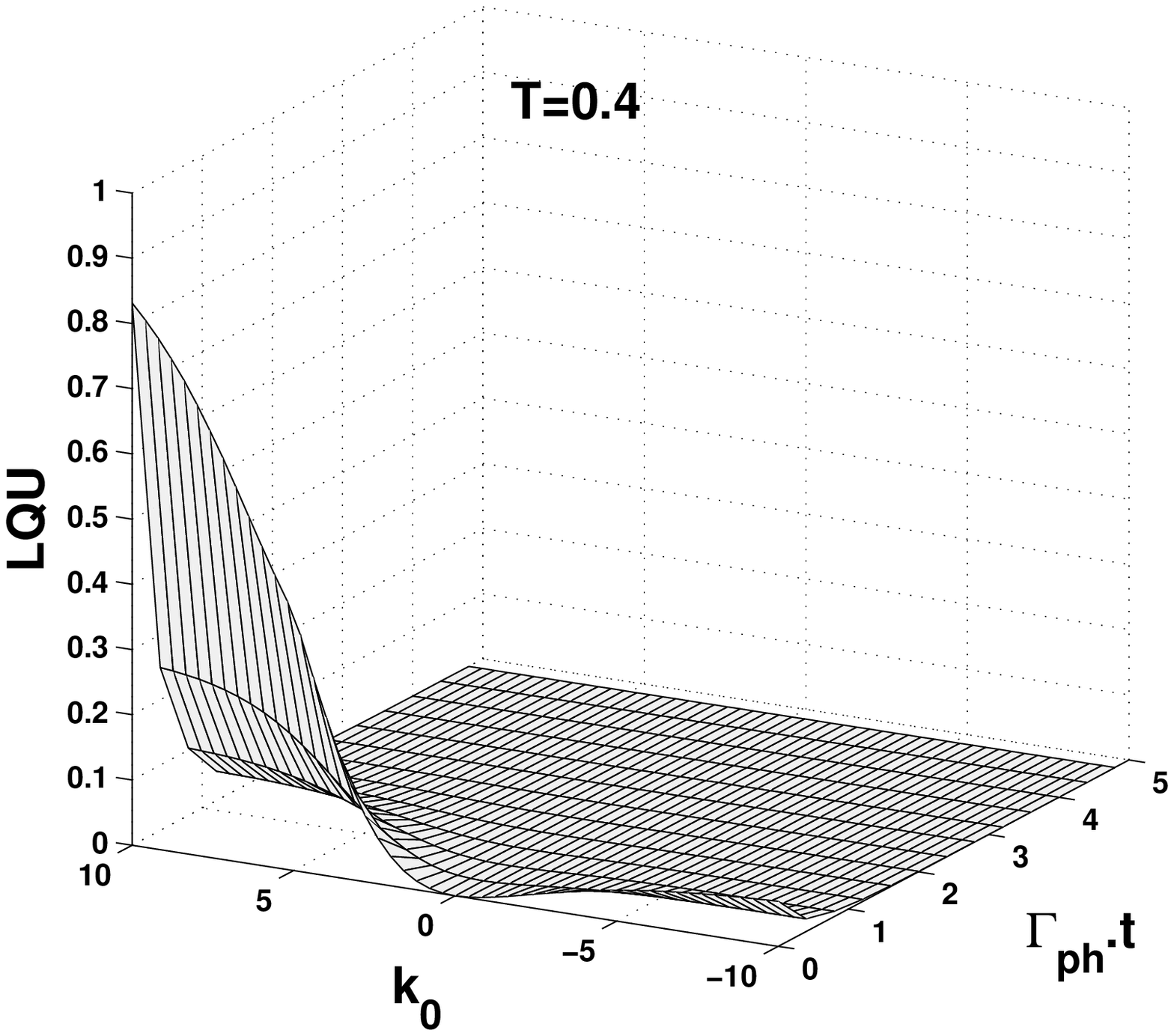}

\caption{QD and LQU vs $\Gamma_{ph}.t$ and $k_0$ in the case of r=1 and T=0.4.}
 \label{fig5}
\end{figure}

\subsection{Amplitude damping channel}
Next we consider time evolution under amplitude noise. Kraus operators for an amplitude damping channel given by $F_1=diag(\sqrt{1-\gamma},1)$ and $F_2=\frac{\sqrt{\gamma}}{2}(\sigma_1-i\sigma_2)$. We find from the appropriate Kraus operators given in \cite{T. Yu} that the following time dependence determine $\rho(t)$ at any time:
\begin{eqnarray} &u(t)=u(0)(1-\gamma)^2,\\\nonumber
&y(t)=y(0)(1-\gamma(t)),\\\nonumber
&w(t)=w(0)(\gamma)^2+u(0)(1-\gamma)\gamma,\\\nonumber
&v(t)=v(0)+u(0)(\gamma)^2+2w(0)\gamma.
\end{eqnarray}
where $\gamma=1-exp(-\Gamma_{am}t)$, and $\Gamma_{am}$ indicate  decay rate of the qubits. The results are in figure 6. We can see from figure 6 that against dephasing channel case, here the behavior of QD and LQU is similar approximately. Both of them decrease asymptotically by growing time, while increase by growing the absolute value of $k_0$.
\begin{figure}
\includegraphics[width=3in]{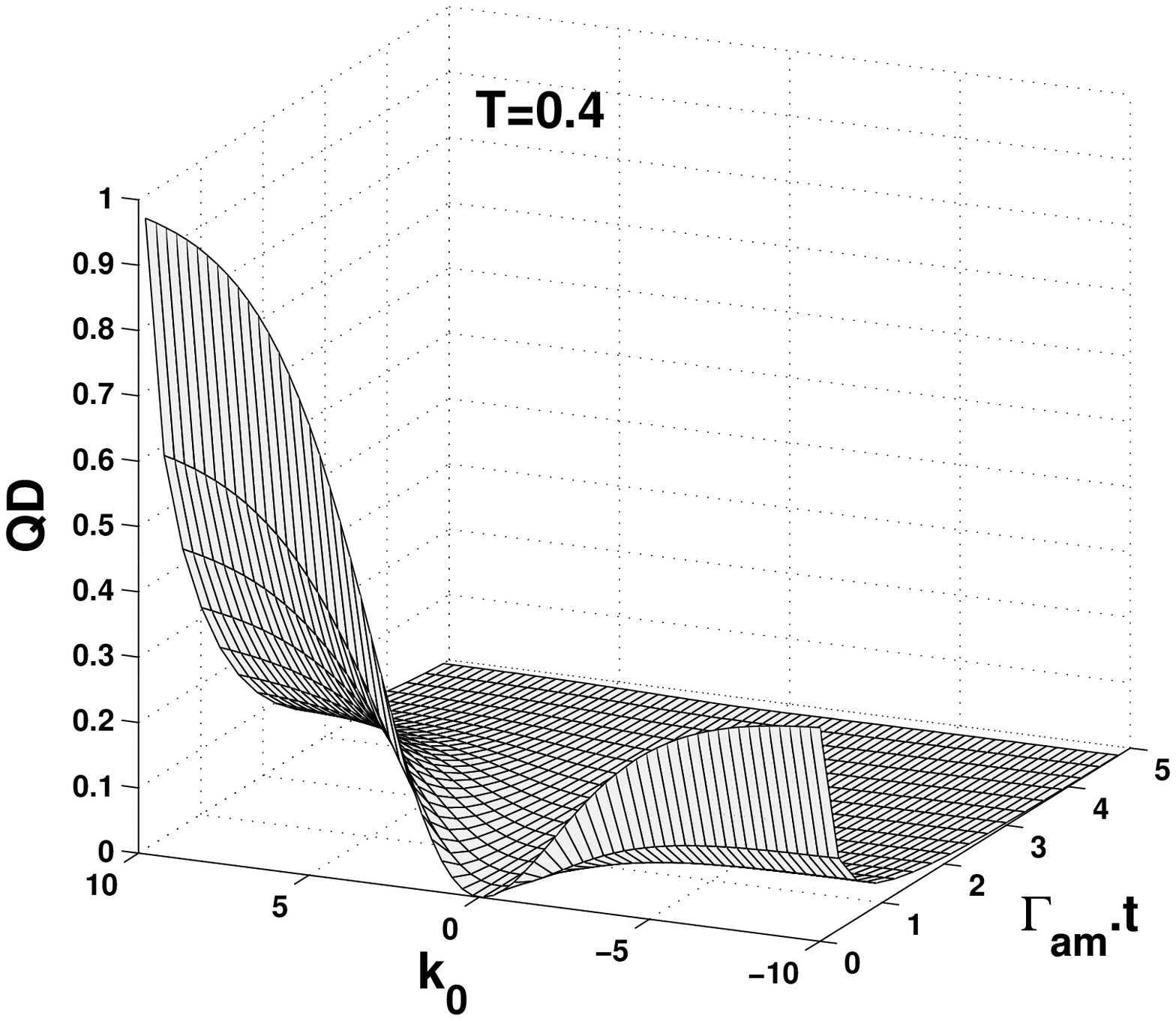}
\includegraphics[width=3in]{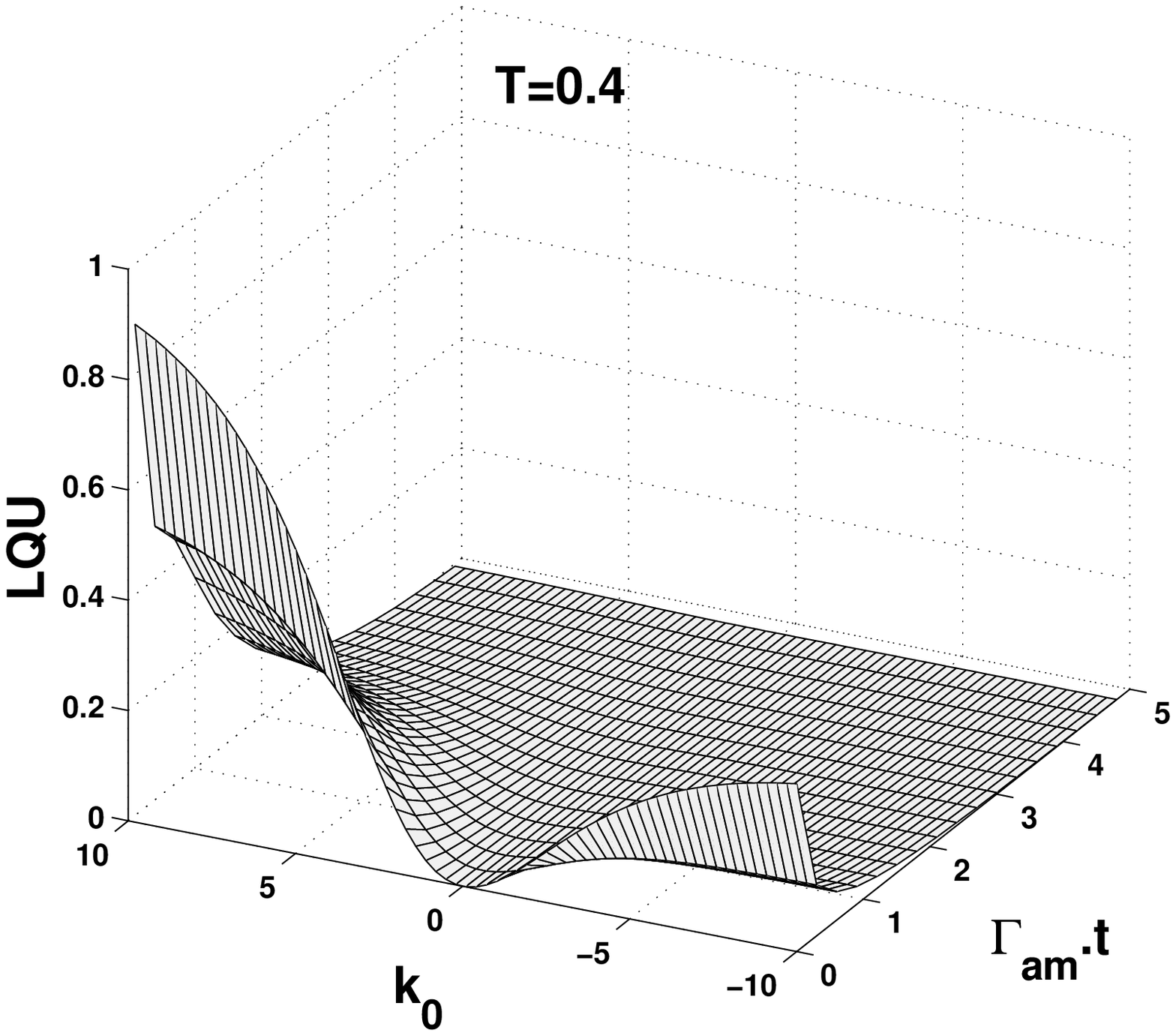}

\caption{QD and LQU vs $\Gamma_{am}.t$ and $k_0$ in the case of r=1 and T=0.4.}
 \label{fig6}
\end{figure}

\section{Conclusion}
In summary, we have investigated QD and LQU in a vertical quantum dot.
Our results imply that the QD and LQU depends on the magnetic field, $k_0$ and the temperature of the system in equilibrium. The behavior of the LQU
is similar to that of the QD  to large extent. They change smoothly without any sudden transitions. The larger $k_0$ is, the larger the QD is and the
stronger the magnetic field is, the smaller the QD
is. With regard to the effect of temperature, we find that the
higher the temperature is, the smaller the QD is. Specifically, the QD is not sensitive to the change of temperature
when the temperature is higher than a value of about 2.   Moreover, we have studied the dynamics
of QD in dephasing and amplitude damping model, and the dynamics of LQU is compared with that of QD. In particular we note the evolution of QD in dephasing channel, which its smoothly behavior at a finite time is noticeable (and remains stable).




\end{document}